\documentclass[
    aps,
    prl,
    reprint,
    superscriptaddress,
    amsmath,
    amssymb,
    longbibliography
]{revtex4-2}

\usepackage[T1]{fontenc}
\usepackage[utf8]{inputenc}
\usepackage{graphicx}
\graphicspath{{figures/}}
\usepackage{bm}
\usepackage{amsmath}
\usepackage{amssymb}
\usepackage{mathtools}
\usepackage{xcolor}
\usepackage{silence}
\usepackage[colorlinks=true,linkcolor=blue,citecolor=blue,urlcolor=blue]{hyperref}
\def\diff{\mathrm{d}}
\newcommand{\rr}{\mathbf{r}}
\newcommand{\kk}{\mathbf{k}}
\newcommand{\spin}{\mathbf{S}}

\def\Dj{\DJ}

\makeatletter
\def\@bibdataout@aps{%
 \immediate\write\@bibdataout{%
  @CONTROL{%
   apsrev42Control%
   \longbibliography@sw{%
    ,author="48",editor="1",pages="0",title="0",year="1"%
   }{%
    ,author="48",editor="1",pages="0",title="",year="1"%
   }%
  }%
 }%
 \if@filesw
  \immediate\write\@auxout{\string\citation{apsrev42Control}}%
 \fi
}%
\makeatother

\begin{document}

\title{Wormhole Geometry from a Magnetic Vortex}

\author{Du\v{s}an \Dj or\dj evi\'{c}}
\affiliation{Faculty of Physics, University of Belgrade, Studentski Trg 12--16, 11000 Belgrade, Serbia}
\author{Fabi\'{a}n Molina}
\affiliation{Departamento de F\'{i}sica, Universidad T\'{e}cnica Federico Santa Mar\'{i}a, Casilla 110, Valpara\'{i}so, Chile}
\author{Vladimir Juri\v{c}i\'{c}}
\thanks{Corresponding author: vladimir.juricic@usm.cl}
\affiliation{Departamento de F\'{i}sica, Universidad T\'{e}cnica Federico Santa Mar\'{i}a, Casilla 110, Valpara\'{i}so, Chile}

\begin{abstract}
Strong coupling to a magnetic texture makes an electron propagate through an emergent curved space. We show that an elementary vortex realizes the exterior spatial geometry of an Ellis wormhole: an ultrastatic throat with radius fixed by the topological charge and Hund exchange, cut off at short distances by the microscopic core. Two separable signatures follow directly: the electron deflection collapses onto a single Ellis curve governed by the vortex winding and exchange coupling, while the spin Berry phase produces a half-flux Aharonov--Bohm response switched on and off by winding parity. The same metric can be emulated in a designer honeycomb lattice, where the valley-symmetrized wave-packet response follows the predicted exterior geodesic. These signatures are accessible through real-space electron deflection and scattering, providing experimentally distinct probes of the emergent geometry and Berry flux. The magnetic vortex thus turns a topological defect into a tunable curved-space lens for electrons in quantum materials and designer lattices.
\end{abstract}

\maketitle

\textit{Introduction.}---Geometry provides a unifying language for electron dynamics in solids: Berry curvature, Berry connections, and the quantum metric control semiclassical motion, transport, and response~\cite{Berry1984,Provost1980,Resta2011,Torma2023}. In slowly varying backgrounds this language can become literal, with low-energy electrons propagating as if they lived in an emergent curved space whose metric is encoded in local velocities, vielbeins, or wave-function geometry. {\color{black}This idea underlies analogue-gravity realizations~\cite{Unruh1981,Barcelo2011}, including curved and strained graphene~\cite{Gonzalez1993,Cortijo2007,Vozmediano2010,deJuan2007,Levy2010,deJuan2012,Wagner2019}, cold-atom and photonic platforms~\cite{Boada2011,Genov2009}, engineered lattices and synthetic dimensions that emulate curved-space Dirac dynamics, horizons, and gravitational lensing through position-dependent couplings~\cite{Konye2023,Konye2022,DeBeule2021,Arguello2024}, and, more recently, quantum-metric lensing by smooth magnetic textures~\cite{Onishi:2025lde}. Magnetic textures thus offer a flexible real-space route to emergent geometry.}

Topological magnetic textures are characterized by discrete winding numbers robust under smooth deformations. They therefore provide a natural organizing principle for emergent electron geometries, suggesting a texture–geometry dictionary in which each topological class selects a distinct effective metric. This raises a sharp question: how does the topology and structure of a magnetic texture determine the emergent geometry seen by electrons? We address this question for one of the elementary topological defects of an ordered magnetic state~\cite{Mermin1979}: the vortex. Around the core, its phase realizes an integer-winding map into the easy-plane order-parameter manifold, \(S^1\), with vortex phase \(\Psi=q\phi\) [Eq.~\eqref{eq:texture}], where \(q\) is the integer winding number and \(\phi\) the azimuthal angle.

Our central result is that this single topological defect converts quantum-metric lensing into the exterior spatial geometry of an Ellis wormhole~\cite{Ellis:1973yv,Bronnikov:1973fh,Morris:1988cz,Visser1995}, whose characteristic scale is fixed by the vortex winding and Hund coupling, $J$. Vortex winding produces an exterior Ellis scale, corresponding in the ideal continuum model to a throat of finite radius, in contrast to the conical geometry of the non-winding spiral~\cite{Onishi:2025lde}. The topological charge fixes the exterior length and the weak-field lensing tail, while the microscopic core profile regularizes the near-throat region. Two signatures follow directly: a universal $q^2/J$ collapse of the deflection, and a parity-locked Berry-flux fingerprint in the scattering. We further show that the same metric can be programmed into an engineered honeycomb lattice, making the vortex--Ellis geometry a target for quantum simulation rather than a peculiarity of one microscopic model. A ubiquitous winding defect thereby maps onto a canonical curved-space lens with two independently tunable responses: the metric scale, set by $q^2/J$, and the Berry flux, set by winding parity. Throughout, ``wormhole'' refers to the emergent spatial geometry seen by the electron: the particle is lensed by the exterior branch of the throat, while the microscopic vortex core supplies the ultraviolet cutoff.

\begin{figure*}[t]
    \centering
    \includegraphics[width=.9\textwidth]{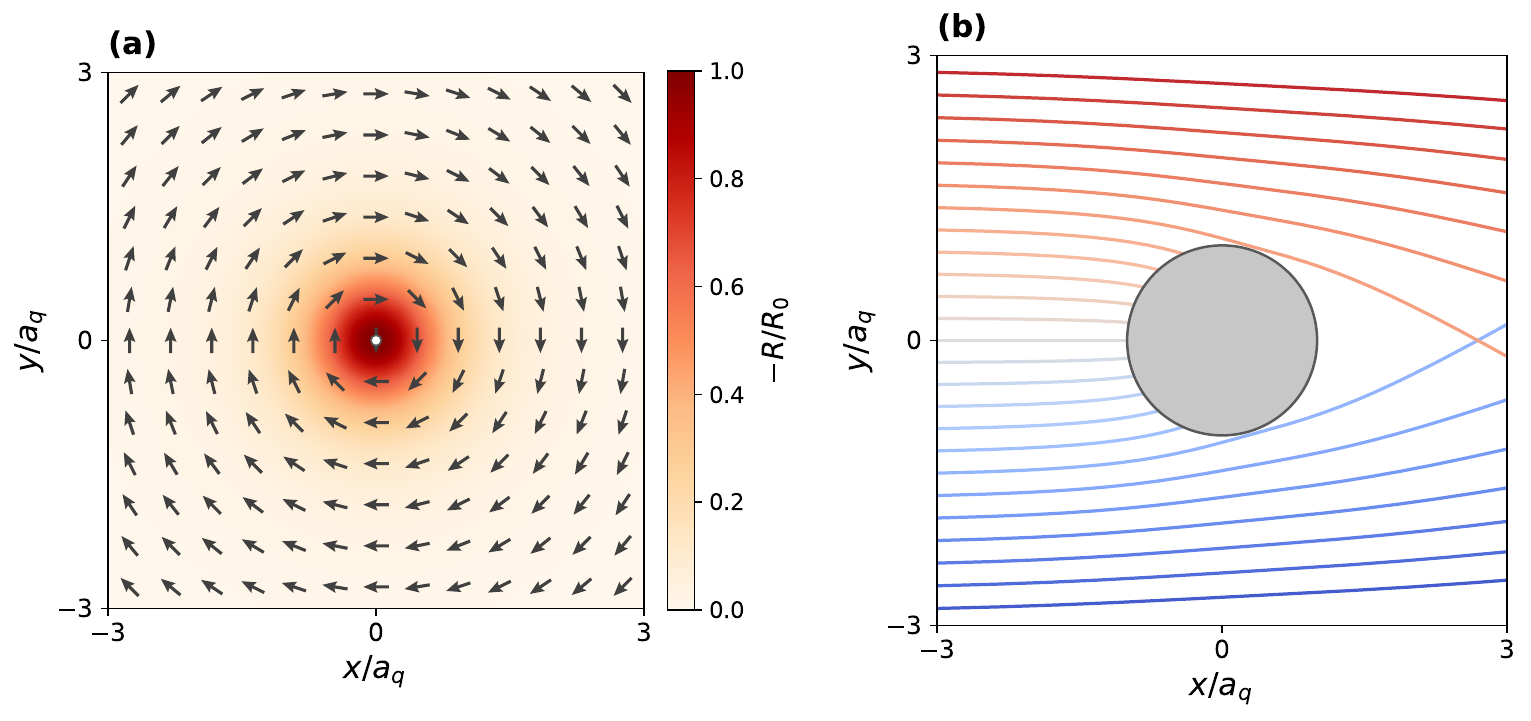}
    \caption{\textbf{Vortex-induced wormhole geometry $(\kk=0)$.}
    (a)~The magnetic vortex texture (arrows: in-plane spin orientation $\spin$)
    {\color{black}with the emergent negative Ricci-curvature magnitude $-R/R_0$ (color), where $R$ is given by Eq.~\eqref{eq:ricci} and $R_0=2/a_q^2$. The magnitude of the negative curvature is largest at the core.}
    (b)~Geodesics of the effective metric~\eqref{eq:polar}, colored by impact parameter defined before Eq.~\eqref{eq:phi_inf}.
    {\color{black} The grey disk  marks the microscopic cutoff of the exterior Ellis geometry.} The exact deflection law and its
    $q^2/J$ collapse are shown in Fig.~\ref{fig:collapse}.}
    \label{fig:vortex}
\end{figure*}

\textit{Emergent metric from the spin texture.}---For an electron whose spin is
locked to a smooth magnetization \(\spin(\rr)\) by a strong Hund exchange coupling, projection onto the locally spin-aligned band gives the quantum-metric
correction~\cite{Onishi:2025lde}
\begin{equation}\label{eq:qmetric}
    g_{ij}
    =
    \delta_{ij}
    +
    \frac{\hbar^2}{4mJ}\,
    \partial_i\spin\cdot\partial_j\spin ,
\end{equation}
with \(m\) the electron mass; see Sec.~S1 of the Supplemental
Material (SM)~\cite{SM}. Thus spatial variations of the magnetization directly
generate departures from the flat metric. {\color{black}The projected description requires strong exchange, $J\gg E$, together with a slowly varying texture~\cite{Onishi:2025lde}, see also Sec.~S1 of the SM~\cite{SM}.} To isolate the vortex contribution,
we consider the planar texture
\begin{equation}\label{eq:texture}
    \spin=\bigl(\cos\Phi,\,-\sin\Phi,\,0\bigr),\,\,{\rm with}\,\,
\Phi(\rr)=\Psi(\rr)+\kk\cdot\rr ,
\end{equation}
where \(\Psi=\arg[(x-X)+i(y-Y)]\) is the vortex phase around the core and
\(\kk\) is a uniform spiral wave vector. Since \(\spin\) is planar,
\(\partial_i\spin\cdot\partial_j\spin=\partial_i\Phi\,\partial_j\Phi\), and
Eq.~\eqref{eq:qmetric} becomes
\begin{equation}\label{eq:decomp}
g_{ij}
=
\delta_{ij}
+
r_0^2\bigl(
\partial_i\Psi\,\partial_j\Psi
+k_i\,\partial_j\Psi
+k_j\,\partial_i\Psi
+k_i k_j
\bigr),
\end{equation}
with a characteristic length scale
\begin{equation}\label{eq:r0}
    r_0=\frac{\hbar}{2\sqrt{mJ}}.
\end{equation}

The four terms separate by range and symmetry. The vortex contribution,
\(\partial_i\Psi\,\partial_j\Psi\sim 1/r^2\), sets the exterior Ellis scale; \(k_i k_j\)
is a constant anisotropic rescaling of the asymptotic metric; and the mixed
terms \(k_i\partial_j\Psi\sim 1/r\) generate direction-dependent far-field
lensing. The exterior Ellis scale is therefore controlled by the vortex sector alone, while
the spiral produces only an anisotropic lensing deformation. We hence set
\(\kk=0\) and focus on the universal vortex geometry; the spiral response is
discussed in Sec.~S2 of the SM~\cite{SM}.

For \(\kk=0\), rotational symmetry gives, in polar coordinates centered on the
vortex core,
\begin{equation}\label{eq:polar}
    \diff s^2=-\diff t^2+\diff r^2+\bigl(r^2+a_q^2\bigr)\diff\phi^2,
\end{equation}
with $a_q=|q|r_0$.  After introducing  \(\rho=(r^2+a_q^2)^{1/2}\), one obtains a more convenient form 
\begin{equation}\label{eq:ellis}
    \diff s^2=-\diff t^2+
    \frac{\diff\rho^2}{1-a_q^2/\rho^2}
    +\rho^2\diff\phi^2 ,
\end{equation}
representing the ultrastatic extension of the equatorial Ellis-wormhole spatial
geometry~\cite{Ellis:1973yv}, with shape function \(b(\rho)=a_q^2/\rho\).
Thus the ideal planar vortex realizes an Ellis spatial metric governed by a
single length scale that grows linearly with \(|q|\). The electron probes only
its ultrastatic exterior branch, \(\rho\ge r_0\), rather than a Lorentzian
traversable wormhole~\cite{Morris:1988cz,Visser1995}. A complementary way to view Eq.~\eqref{eq:ellis}
is as the 2+1-dimensional Ellis geometry in its ultrastatic
form: the redshift function multiplying $dt^2$ is constant,
so all nontrivial geometry resides in the spatial metric. In the magnetic system this geometry describes the continuum exterior of the throat: the microscopic core supplies the short-distance cutoff, and no second asymptotic region is physically traversed.

Equation~\eqref{eq:polar} is the continuum exterior metric of an ideal easy-plane vortex; a core-regularized texture reads
\begin{equation}\label{eq:core_texture}
    \spin(r,\phi)=
    \bigl(\sin\theta(r)\cos q\phi,
    -\sin\theta(r)\sin q\phi,
    \cos\theta(r)\bigr),
\end{equation}
which gives
\begin{equation}\label{eq:core_metric}
    g_{\phi\phi}=r^2+r_0^2 q^2\sin^2\theta(r),
    \qquad
    g_{rr}=1+r_0^2[\theta'(r)]^2 .
\end{equation}
The microscopic core smoothly terminates the geometry at the origin, while the radial correction remains short ranged and disappears in the exterior region. For \(r\gg\xi_{\rm core}\), however, \(\sin\theta(r)\simeq1\) and \(\theta'(r)\simeq0\), so Eq.~\eqref{eq:polar} and the weak-field lensing laws below are recovered. The projected electron therefore realizes the exterior, core-truncated Ellis geometry, with the derivation given in Sec.~S1 of the SM~\cite{SM}.

The Ricci scalar is
\begin{equation}\label{eq:ricci}
    R=-\frac{2a_q^2}{(r^2+a_q^2)^2}
     =-\frac{2a_q^2}{\rho^4}
     =-\frac{8\hbar^2 mJ\,q^2}
     {(\hbar^2 q^2+4mJr^2)^2}.
\end{equation}
The curvature is negative everywhere, {\color{black}reaching its largest magnitude} \(R=-2/a_q^2\) at the throat, and
decays as \(\rho^{-4}\), so the geometry is asymptotically flat. Its peak value,
fixed by the winding through \(a_q\), is a direct geometric imprint of the vortex
charge. As the leading term of the \(J^{-1}\) expansion~\cite{Onishi:2025lde},
Eq.~\eqref{eq:qmetric} controls the continuum geometry in the far field, with the
microscopic core regularizing the throat; see
Sec.~S1 of the SM~\cite{SM}.

\textit{Lensing law and universal scaling.}---Electron deflection probes the vortex geometry most directly. For unit-speed geodesics of the metric~\eqref{eq:polar} the conserved angular momentum is \(L=(r^2+a_q^2)\dot\phi\), so the impact parameter is \(b=L\) and a trajectory with \(b>a_q\) turns at \(r_{\min}=\sqrt{b^2-a_q^2}\). The accumulated angle is
\begin{equation}\label{eq:phi_inf}
    \varphi_\infty(b)
    =
    \int_0^1
    \frac{\diff t}
    {\sqrt{1-t^2}\sqrt{1-(a_q/b)^2t^2}}
    =
    K_{\mathrm{ell}}\!\left(\frac{a_q}{b}\right),
\end{equation}
as derived in Sec.~S3 of the SM~\cite{SM}. The total
deflection is therefore
\begin{equation}\label{eq:lensing}
    \Theta(b)
    =
    2K_{\mathrm{ell}}\!\left(\frac{a_q}{b}\right)-\pi,
    \qquad b>a_q ,
\end{equation}
coinciding with the one-sided Ellis-wormhole lensing law
\cite{Nakajima:2012pu,ChetouaniClement1984}. As \(b\to a_q^+\), the elliptic
integral diverges logarithmically, the strong-deflection limit of the ultrastatic Ellis geometry~\cite{Tsukamoto2016}: geodesics wind about the throat
[Fig.~\ref{fig:vortex}(b)]. In the magnetic system this regime is regularized
by the finite vortex core, where the continuum \(J^{-1}\) expansion reaches its
short-distance cutoff.

\begin{figure}[t]
    \centering
    \includegraphics[width=\columnwidth]{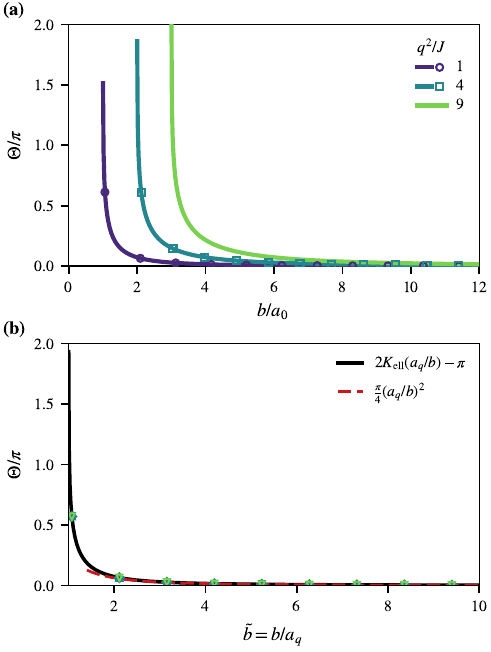}
    \caption{\textbf{Universal $q^2/J$ scaling.} (a)~Deflection angle $\Theta$ for three values of $q^2/J$, with $b$ in units of the fixed length $a_0=\hbar/(2\sqrt{mJ_0})$. Because $a_q^2=q^2\hbar^2/(4mJ)$, configurations with the same $q^2/J$ have the same $a_q$ and therefore the same deflection without rescaling. {\color{black}For each color, the solid curve corresponds to a configuration at \(J=J_0\),
while the open marker denotes a distinct configuration at \(J=4J_0\) with the
same \(q^2/J\), and hence the same deflection without rescaling:
\((q,J)=(1,J_0)\) is paired with \((2,4J_0)\), and
\((q,J)=(2,J_0)\) with \((4,4J_0)\).} (b)~Measured in each configuration's own Ellis scale, $\tilde b=b/a_q$, the data collapse onto Eq.~\eqref{eq:lensing} (solid), with the weak-field tail $(\pi/4)\tilde b^{-2}$ (dashed).}
    \label{fig:collapse}
\end{figure}

For \(b\gg a_q\), the far-field expansion \(K_{\mathrm{ell}}(x)=\tfrac{\pi}{2}(1+x^2/4+\cdots)\) gives
\begin{equation}\label{eq:weak}
    \Theta(b)
    \simeq
    \frac{\pi a_q^2}{4b^2}
    =
    \frac{\pi q^2\hbar^2}{16mJ\,b^2}.
\end{equation}
The weak-deflection tail is universal and set solely by the Ellis scale
\(a_q\), so the lensing curves collapse when plotted against \(b/a_q\)
[Fig.~\ref{fig:collapse}]. Since the winding number and exchange coupling enter
only through \(a_q\), configurations with the same Ellis scale exhibit
identical deflection already in laboratory units
[Fig.~\ref{fig:collapse}(a)], before any rescaling. See also Sec.~S3 of the
SM~\cite{SM}.

The non-constant trace of the quantum metric also generates a scalar potential
for the projected electron,
$U_{\mathrm{geom}}(r)=\hbar^2q^2/(8mr^2)$, which is long ranged and must be
distinguished from the finite-$J$ metric lensing. It enters the scattering
problem through Eq.~\eqref{eq:radial_full} below, whereas the geodesic
deflection~\eqref{eq:lensing}--\eqref{eq:weak} follows from the
metric~\eqref{eq:polar} alone. In the far field the eikonal requirement is
local, $\kappa L_R(b)\gg1$, with
$L_R=|R|^{-1/2}\simeq b^2/(\sqrt{2}a_q)$ for $b\gg a_q$; it can therefore
coexist with the strong-exchange condition
$\kappa a_q=|q|\sqrt{E/(2J)}\ll1$ for an elementary vortex. The two projected
contributions also have different control parameters:
$a_q^2\propto q^2/J$, whereas $U_{\mathrm{geom}}\propto q^2$ is independent of
$J$ at fixed winding. Varying $J$ therefore separates metric lensing from the
scalar contribution. Details are given in Secs.~S1, S4, and S5 of the
SM~\cite{SM}.

{\color{black}
\textit{Geometric scattering and Berry phase.}---Scattering provides a complementary probe of the same emergent geometry. In addition to the metric, projection onto the spin-polarized band generates a Berry connection, the texture analogue of emergent electromagnetism~\cite{Volovik1987}. For an easy-plane vortex, \(\oint\mathcal A=\pi q\), corresponding to the effective Aharonov--Bohm flux \(\gamma=q/2\), and therefore shifting angular momentum as \(\ell\to\ell-\gamma\)~\cite{AharonovCasher1984,AharonovBohm1959}. Accordingly, for a partial wave \(\psi_\ell=R_\ell(r)e^{i\ell\phi}\), the  Laplace--Beltrami operator of metric~\eqref{eq:polar}, together with the Berry shift and the geometric scalar potential derived in Sec.~S4 of the SM~\cite{SM}, yields
\begin{equation}\label{eq:radial_full}
R_\ell''+\frac{r}{r^2+a_q^2}R_\ell'+\left[\kappa^2-\frac{(\ell-\gamma)^2}{r^2+a_q^2}-\frac{2m}{\hbar^2}U_{\mathrm{geom}}\right]R_\ell=0,
\end{equation}
with \(E=\hbar^2\kappa^2/2m\). In the semiclassical large-\(|\ell|\) limit, the metric phase shift follows from \(\Theta=-2\,\diff\delta_\ell/\diff\ell\),
\[
\delta_\ell^{\mathrm{metric}}\simeq\frac{\pi(\kappa a_q)^2}{8|\ell-\gamma|},
\]
and hence carries the same geometric scale \(a_q^2\propto q^2/J\) as the weak-field deflection in Eq.~\eqref{eq:weak}.

By contrast, the Berry contribution is fixed by the enclosed flux and gives the exact Aharonov--Bohm cross section
\begin{equation}\label{eq:abxsec}
\left.\frac{\diff\sigma}{\diff\varphi}\right|_{\mathrm{AB}}=\frac{\sin^2(\pi\gamma)}{2\pi\kappa\,\sin^2(\varphi/2)}.
\end{equation}
Since \(\gamma=q/2\), this fingerprint is maximal for odd winding and absent for even winding. The finite-throat metric and scalar-potential phase shifts add diffraction structure to this universal background. Their partial-wave construction, convergence benchmarks, and diagnostic cross section are presented in Sec.~S5 of the SM~\cite{SM}. 
}

\begin{figure*}[t]
    \centering
    \includegraphics[width=\textwidth]{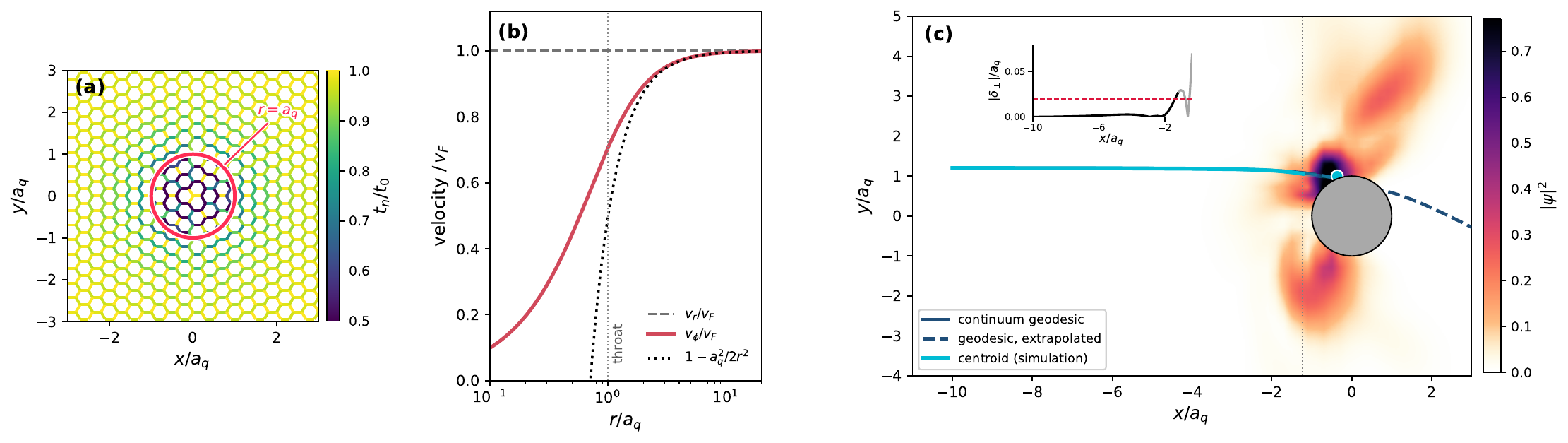}
   \caption{{\color{black}\textbf{Designer honeycomb emulator.} (a)~Engineered nearest-neighbor hoppings (color: $t_n/t_0$) realizing the inverse vielbein of the core-truncated Ellis metric to leading order, with tangential bonds suppressed near the core [Eq.~\eqref{eq:hop}]. (b)~Target velocity profile, $v_r=v_F$ and $v_\phi/v_F=r/\sqrt{r^2+a_q^2}$. (c)~Valley-symmetrized wave-packet evolution (Supplemental Movie~1): averaging otherwise identical $K$ and $K'$ packets removes the valley-odd pseudogauge deflection and isolates the valley-even metric response. In the controlled exterior region the centroid follows the core-regularized continuum geodesic over $8.6\,a_q$ of path to $|\delta_\perp|<0.02a_q$; the dashed curve extrapolates the geodesic beyond this tracking window.}}
\label{fig:sim}
\end{figure*}

{\color{black}\textit{Designer honeycomb emulator.}---}The magnetic vortex is one route to this core-truncated Ellis geometry; an alternative is to program the Dirac velocity tensor so that its vielbein reproduces Eq.~\eqref{eq:ellis}. A honeycomb lattice with engineered hoppings provides a direct {\color{black}effective-metric implementation in the controlled exterior region}, in the spirit of curved-space lattice emulators that reproduce fermion geodesics through spatially varying couplings~\cite{Konye2022,Konye2023}.
For the orthonormal frame \(e^1=\diff r\), \(e^2=f(r)\diff\phi\), with
\(f(r)=\sqrt{r^2+a_q^2}\), the inverse vielbein fixes the angular velocity,
\begin{equation}\label{eq:vel}
    \frac{v_\phi(r)}{v_F}
    =
    \frac{r}{\sqrt{r^2+a_q^2}}
    \simeq
    1-\frac{a_q^2}{2r^2},
\end{equation}
while the radial velocity remains \(v_F\). This velocity profile is obtained by
suppressing the tangential hopping near the core,
\begin{equation}\label{eq:hop}
    t_n(\mathbf R)
    =
    t_0\left[
    1-\frac{a_q^2}{2(r^2+\xi^2)}
    (\hat{\bm\delta}_n\cdot\hat{\mathbf e}_\phi)^2
    \right],
\end{equation}
where \(\xi\) regularizes the microscopic core [Fig.~\ref{fig:sim}(a,b)]. {\color{black}Artificial graphene~\cite{Polini2013}, photonic~\cite{Ozawa2019}, and topolectric platforms~\cite{Lee2018} offer independent control of local couplings. The bond modulation in Eq.~\eqref{eq:hop} reproduces the Ellis velocity profile to leading order in the exterior, while its spatial variation also induces a valley-odd pseudogauge contribution. The valley degree of freedom is specific to the honeycomb emulator and is absent from the original magnetic problem. To this end, we average over the two valleys: the pseudogauge contribution cancels, whereas the common metric deflection is retained, thereby isolating the curved-space dynamics encoded by the original vortex problem.} The construction therefore realizes a programmable metric emulator rather than a literal magnetic vortex; details and geodesic-level checks are given in Sec.~S6 of the SM~\cite{SM}. {\color{black}In the wave-packet calculation we set the independently tunable on-site scalar term to zero, thereby isolating the metric response.}

{\color{black}A real-time tight-binding simulation provides a direct dynamical test of the engineered metric. We propagate otherwise identical packets near the two valleys, with the Dirac pseudospinor in each valley chosen for initial propagation along $+x$, and average their densities and centroids. This valley symmetrization cancels the valley-odd pseudogauge bending generated by the bond modulation and isolates the valley-even metric response, which we compare with the geodesic of the same core-regularized continuum metric.} {\color{black}Within the controlled exterior tracking window, the averaged centroid bends toward the core and follows the core-regularized geodesic over $8.6\,a_q$ of path with transverse deviation}
\(|\delta_\perp|<0.02\,a_q\) [Fig.~\ref{fig:sim}(c), Sec.~S6, and Supplemental
Movie~1]. {\color{black}This agreement shows that, after removing the valley-odd pseudogauge component, the engineered hopping profile reproduces the exterior vielbein to the quoted accuracy rather than only the sign of the deflection.} {\color{black}We deliberately restrict this comparison to the exterior tracking window, where the packet remains ray-like and boundary effects are negligible. }

{\color{black}
The magnetic and designer realizations involve different microscopic controls. In a magnetic system, the Ellis scale follows directly from the projected electron parameters: for $m\sim0.4m_e$ and $J\sim0.1$--$1\,{\rm eV}$, one obtains $a_q\simeq0.69$--$0.22\,{\rm nm}$ for $q=1$. In a designer honeycomb emulator, by contrast, $a_q$ is programmed directly through the hopping profile. The simulation of Fig.~\ref{fig:sim} uses {\color{black}$a_q=19$} lattice constants and $\xi=\sqrt{0.12}\,a_q\simeq0.35a_q$ (Sec.~S6 of SM~\cite{SM}). The far-field deflection at $b\simeq5$--$10a_q$ is core-insensitive, while resolving the near-throat region requires $\xi\ll a_q$, as quantified in Fig.~S2 of the SM~\cite{SM}. For the same impact parameters, Eq.~\eqref{eq:weak} gives $\Theta\simeq0.45^\circ$--$1.8^\circ$, corresponding to a transverse shift of order $0.1$--$0.6\,{\rm nm}$ over a $\sim20\,{\rm nm}$ path for the above magnetic scale. The most characteristic discriminants are thus the vortex-centered deflection, its $q^2/J$ scaling, and the independent odd--even Berry-flux response. Further estimates and numerical controls are given in Sec.~S6 of the SM~\cite{SM}.
}

\textit{Discussion.}---{\color{black}We have shown that a single magnetic vortex imprints the exterior, core-truncated spatial geometry of an Ellis wormhole on a Hund-coupled itinerant electron. The Ellis length is fixed by the vortex winding and exchange coupling, $a_q=|q|\hbar/(2\sqrt{mJ})$, leading to a closed-form lensing law and a universal $q^2/J$ collapse of the far-field deflection. The Berry connection supplies an independent scattering fingerprint, maximal for odd winding and absent for even. A designer honeycomb lattice provides a complementary emulator of the same exterior vielbein, with the valley-symmetrized packet dynamics reproducing its geodesic response in the controlled region. The result therefore ties a canonical curved-space geometry to a topological defect ubiquitous in easy-plane magnets and magnetic thin films~\cite{Papanicolaou1991,Shinjo2000,Wachowiak2002,NagaosaTokura2013,Gobel2021}.}

The construction also suggests several immediate extensions. A nonzero spiral
wave vector \(\kk\) leaves the exterior Ellis scale intact but adds a long-range,
direction-dependent lensing deformation through the mixed term in
Eq.~\eqref{eq:decomp}. Vortex--antivortex configurations should realize
screened two-center lenses, while the marginal \(1/r^2\) scalar potential near
the throat may generate defect-localized resonances visible in local spectral
probes. These extensions are experimentally accessible through controlled
magnetic-texture engineering, real-space electron deflection, and local
spectroscopy, providing complementary probes of the emergent geometry.

More broadly, our result points to a texture--geometry dictionary for magnetic
backgrounds, in which different topological sectors can select distinct
families of effective geometries for the projected electron. Two entries are
already apparent: the radial spiral studied in
Ref.~\cite{Onishi:2025lde} realizes a conical, point-mass-like geometry,
whereas the azimuthal vortex studied here realizes an exterior,
core-truncated Ellis lens. Extending this dictionary to merons, skyrmions, and
multi-defect textures~\cite{NagaosaTokura2013,Gobel2021}, whose emergent
electrodynamics is already well established~\cite{Schulz2012}, would make
magnetic order a versatile route to engineering curved-space electron
dynamics. The magnetic vortex thus provides a tunable building block for
curved-space electron dynamics in quantum materials and programmable lattices.

\begin{acknowledgments}
\textit{Acknowledgments.}---{D.\Dj. and V.J. acknowledge the funding provided by the Faculty of Physics of the University of Belgrade, through a grant number 451-03-137/2025-03/200162 from the Ministry of Education, Science, and Technological Development of the Republic of Serbia and by the Science Fund of the Republic of Serbia, Serbian Scientific Cooperation Program with the Diaspora: Support for Visits of Diaspora Scientists, under Project TopArtGravity No. 220.} V. J. acknowledges the support by Fondecyt (Chile) Grant No. 1230933. 
\end{acknowledgments}

\textit{Data availability.}---The codes that generate all figures and movies  are
openly available at Ref.~\cite{Code}. Anthropic's Claude Opus 5 assisted with the figures, numerical checks,
and manuscript preparation.

\bibliography{ref}

@article{Morris:1988cz,
    author = "Morris, M. S. and Thorne, K. S.",
    title = "{Wormholes in space-time and their use for interstellar travel: A tool for teaching general relativity}",
    doi = "10.1119/1.15620",
    journal = "Am. J. Phys.",
    volume = "56",
    pages = "395--412",
    year = "1988"
}

@article{Ellis:1973yv,
    author = "Ellis, H. G.",
    title = "{Ether flow through a drainhole - a particle model in general relativity}",
    doi = "10.1063/1.1666161",
    journal = "J. Math. Phys.",
    volume = "14",
    pages = "104--118",
    year = "1973"
}

@article{Onishi:2025lde,
    author = "Onishi, Yugo and Paul, Nisarga and Fu, Liang",
    title = "{Emergent curved space and gravitational lensing in quantum materials}",
    eprint = "2506.04335",
    archivePrefix = "arXiv",
    primaryClass = "cond-mat.mes-hall",
    doi = "10.1103/qxnw-8q4y",
    journal = "Phys. Rev. B",
    volume = "113",
    number = "2",
    pages = "024401",
    year = "2026"
}

@article{Nakajima:2012pu,
    author = "Nakajima, Koki and Asada, Hideki",
    title = "{Deflection angle of light in an Ellis wormhole geometry}",
    eprint = "1204.3710",
    archivePrefix = "arXiv",
    primaryClass = "gr-qc",
    doi = "10.1103/PhysRevD.85.107501",
    journal = "Phys. Rev. D",
    volume = "85",
    pages = "107501",
    year = "2012"
}

@article{Gonzalez1993,
author = {Gonz{\'a}lez, J. and Guinea, F. and Vozmediano, M. A. H.},
title = {The electronic spectrum of fullerenes from the Dirac equation},
journal = {Nuclear Physics B},
volume = {406},
pages = {771--794},
year = {1993},
doi = {10.1016/0550-3213(93)90013-3},
eprint = {cond-mat/9208004},
archivePrefix = {arXiv}
}

@article{Cortijo2007,
author = {Cortijo, Alberto and Vozmediano, Mar{\'i}a A. H.},
title = {Electronic properties of curved graphene sheets},
journal = {Europhysics Letters},
volume = {77},
pages = {47002},
year = {2007},
doi = {10.1209/0295-5075/77/47002},
eprint = {cond-mat/0603717},
archivePrefix = {arXiv}
}

@article{deJuan2012,
author = {de Juan, Fernando and Sturla, Mauricio and Vozmediano, Mar{\'i}a A. H.},
title = {Space dependent Fermi velocity in strained graphene},
journal = {Physical Review Letters},
volume = {108},
pages = {227205},
year = {2012},
doi = {10.1103/PhysRevLett.108.227205},
eprint = {1201.2656},
archivePrefix = {arXiv},
primaryClass = {cond-mat.mes-hall}
}

@article{Berry1984,
author = {Berry, M. V.},
title = {Quantal phase factors accompanying adiabatic changes},
journal = {Proceedings of the Royal Society of London A},
volume = {392},
pages = {45--57},
year = {1984},
doi = {10.1098/rspa.1984.0023}
}

@article{Provost1980,
author = {Provost, J. P. and Vallee, G.},
title = {Riemannian structure on manifolds of quantum states},
journal = {Communications in Mathematical Physics},
volume = {76},
pages = {289--301},
year = {1980},
doi = {10.1007/BF02193559}
}

@article{Resta2011,
author = {Resta, Raffaele},
title = {The insulating state of matter: a geometrical theory},
journal = {European Physical Journal B},
volume = {79},
pages = {121--137},
year = {2011},
doi = {10.1140/epjb/e2010-10874-4}
}

@article{Torma2023,
author = {T{\"o}rm{\"a}, P{\"a}ivi},
title = {Essay: Where Can Quantum Geometry Lead Us?},
journal = {Physical Review Letters},
volume = {131},
pages = {240001},
year = {2023},
doi = {10.1103/PhysRevLett.131.240001}
}

@article{Mermin1979,
author = {Mermin, N. D.},
title = {The topological theory of defects in ordered media},
journal = {Reviews of Modern Physics},
volume = {51},
pages = {591--648},
year = {1979},
doi = {10.1103/RevModPhys.51.591}
}

@article{Unruh1981,
author = {Unruh, W. G.},
title = {Experimental black-hole evaporation?},
journal = {Physical Review Letters},
volume = {46},
pages = {1351--1353},
year = {1981},
doi = {10.1103/PhysRevLett.46.1351}
}

@article{Barcelo2011,
author = {Barcel{\'o}, Carlos and Liberati, Stefano and Visser, Matt},
title = {Analogue gravity},
journal = {Living Reviews in Relativity},
volume = {14},
pages = {3},
year = {2011},
doi = {10.12942/lrr-2011-3}
}

@article{Vozmediano2010,
author = {Vozmediano, M. A. H. and Katsnelson, M. I. and Guinea, F.},
title = {Gauge fields in graphene},
journal = {Physics Reports},
volume = {496},
pages = {109--148},
year = {2010},
doi = {10.1016/j.physrep.2010.07.003}
}

@article{deJuan2007,
author = {de Juan, Fernando and Cortijo, Alberto and Vozmediano, Mar{\'i}a A. H.},
title = {Charge inhomogeneities due to smooth ripples in graphene sheets},
journal = {Physical Review B},
volume = {76},
pages = {165409},
year = {2007},
doi = {10.1103/PhysRevB.76.165409}
}

@article{Levy2010,
author = {Levy, N. and Burke, S. A. and Meaker, K. L. and Panlasigui, M. and Zettl, A. and Guinea, F. and Castro Neto, A. H. and Crommie, M. F.},
title = {Strain-induced pseudo-magnetic fields greater than 300 tesla in graphene nanobubbles},
journal = {Science},
volume = {329},
pages = {544--547},
year = {2010},
doi = {10.1126/science.1191700}
}

@article{Boada2011,
author = {Boada, O. and Celi, A. and Latorre, J. I. and Lewenstein, M.},
title = {Dirac equation for cold atoms in artificial curved spacetimes},
journal = {New Journal of Physics},
volume = {13},
pages = {035002},
year = {2011},
doi = {10.1088/1367-2630/13/3/035002}
}

@article{Genov2009,
author = {Genov, Dentcho A. and Zhang, Shuang and Zhang, Xiang},
title = {Mimicking celestial mechanics in metamaterials},
journal = {Nature Physics},
volume = {5},
pages = {687--692},
year = {2009},
doi = {10.1038/nphys1338}
}

@article{Bronnikov:1973fh,
author = {Bronnikov, K. A.},
title = {Scalar-tensor theory and scalar charge},
journal = {Acta Physica Polonica B},
volume = {4},
pages = {251--266},
year = {1973}
}

@book{Visser1995,
author = {Visser, Matt},
title = {Lorentzian Wormholes: From Einstein to Hawking},
publisher = {AIP Press},
address = {Woodbury, NY},
year = {1995}
}

@article{ChetouaniClement1984,
author = {Chetouani, L. and Cl{\'e}ment, G.},
title = {Geometrical optics in the Ellis geometry},
journal = {General Relativity and Gravitation},
volume = {16},
pages = {111--119},
year = {1984},
doi = {10.1007/BF00759650}
}

@article{AharonovCasher1984,
author = {Aharonov, Y. and Casher, A.},
title = {Topological quantum effects for neutral particles},
journal = {Physical Review Letters},
volume = {53},
pages = {319--321},
year = {1984},
doi = {10.1103/PhysRevLett.53.319}
}

@article{AharonovBohm1959,
author = {Aharonov, Y. and Bohm, D.},
title = {Significance of electromagnetic potentials in the quantum theory},
journal = {Physical Review},
volume = {115},
pages = {485--491},
year = {1959},
doi = {10.1103/PhysRev.115.485}
}

@article{Gobel2021,
author = {G{\"o}bel, B. and Mertig, I. and Tretiakov, O. A.},
title = {Beyond skyrmions: Review and perspectives of alternative magnetic quasiparticles},
journal = {Physics Reports},
volume = {895},
pages = {1--28},
year = {2021},
doi = {10.1016/j.physrep.2020.10.001}
}

@article{Schulz2012,
author = {Schulz, T. and Ritz, R. and Bauer, A. and Halder, M. and Wagner, M. and Franz, C. and Pfleiderer, C. and Everschor, K. and Garst, M. and Rosch, A.},
title = {Emergent electrodynamics of skyrmions in a chiral magnet},
journal = {Nature Physics},
volume = {8},
pages = {301--304},
year = {2012},
doi = {10.1038/nphys2231}
}

@misc{SM,
year = {2026},
note = {See Supplemental Material for Sec.~S1, the emergent metric and wormhole curvature; Sec.~S2, the role of the spiral wave vector; Sec.~S3, the elliptic lensing integral; Sec.~S4, Berry flux and scalar potential; Sec.~S5, the partial-wave cross section with Aharonov--Bohm resummation; Sec.~S6, the tight-binding construction and wave-packet simulation; and Supplemental Movie~1}
}

@article{Wagner2019,
author = {Wagner, Glenn and de Juan, Fernando and Nguyen, Dung X.},
title = {Landau levels in curved space realized in strained graphene},
journal = {SciPost Physics Core},
volume = {5},
pages = {029},
year = {2022},
doi = {10.21468/SciPostPhysCore.5.2.029},
eprint = {1911.02028},
archivePrefix = {arXiv},
primaryClass = {cond-mat.mes-hall}
}

@article{Konye2023,
  title = {Anisotropic optics and gravitational lensing of tilted {W}eyl fermions},
  author = {K\"onye, Vikt\'or and Mertens, Lotte and Morice, Corentin and Chernyavsky, Dmitry and Moghaddam, Ali G. and van Wezel, Jasper and van den Brink, Jeroen},
  journal = {Phys. Rev. B},
  volume = {107},
  pages = {L201406},
  year = {2023},
  doi = {10.1103/PhysRevB.107.L201406}
}

@article{Konye2022,
  title = {Horizon physics of quasi-one-dimensional tilted {W}eyl cones on a lattice},
  author = {K\"onye, Vikt\'or and Morice, Corentin and Chernyavsky, Dmitry and Moghaddam, Ali G. and van den Brink, Jeroen and van Wezel, Jasper},
  journal = {Phys. Rev. Research},
  volume = {4},
  pages = {033237},
  year = {2022},
  doi = {10.1103/PhysRevResearch.4.033237}
}

@article{DeBeule2021,
  title = {Artificial event horizons in {W}eyl semimetal heterostructures and their non-equilibrium signatures},
  author = {De Beule, Christophe and Groenendijk, Solofo and Meng, Tobias and Schmidt, Thomas L.},
  journal = {SciPost Phys.},
  volume = {11},
  pages = {095},
  year = {2021},
  doi = {10.21468/SciPostPhys.11.5.095}
}

@article{Arguello2024,
  title = {Synthetic dimensions for topological and quantum phases},
  author = {Arg\"uello-Luengo, Javier and Bhattacharya, Utso and Celi, Alessio and Chhajlany, Ravindra W. and Grass, Tobias and Plodzien, Marcin and Rakshit, Debraj and Salamon, Tymoteusz and Stornati, Paolo and Tarruell, Leticia and Lewenstein, Maciej},
  journal = {Commun. Phys.},
  volume = {7},
  pages = {143},
  year = {2024},
  doi = {10.1038/s42005-024-01636-3}
}

@article{Volovik1987,
  title = {Linear momentum in ferromagnets},
  author = {Volovik, G. E.},
  journal = {J. Phys. C: Solid State Phys.},
  volume = {20},
  number = {7},
  pages = {L83},
  year = {1987},
  doi = {10.1088/0022-3719/20/7/003}
}

@article{Tsukamoto2016,
  title = {Strong deflection limit analysis and gravitational lensing of an {E}llis wormhole},
  author = {Tsukamoto, Naoki},
  journal = {Phys. Rev. D},
  volume = {94},
  pages = {124001},
  year = {2016},
  doi = {10.1103/PhysRevD.94.124001}
}

@article{Polini2013,
  title = {Artificial honeycomb lattices for electrons, atoms and photons},
  author = {Polini, Marco and Guinea, Francisco and Lewenstein, Maciej and Manoharan, Hari C. and Pellegrini, Vittorio},
  journal = {Nat. Nanotechnol.},
  volume = {8},
  pages = {625},
  year = {2013},
  doi = {10.1038/nnano.2013.161}
}

@article{Lee2018,
  title = {Topolectrical circuits},
  author = {Lee, Ching Hua and Imhof, Stefan and Berger, Christian and Bayer, Florian and Brehm, Johannes and Molenkamp, Laurens W. and Kiessling, Tobias and Thomale, Ronny},
  journal = {Commun. Phys.}, volume = {1}, pages = {39}, year = {2018},
  doi = {10.1038/s42005-018-0035-2}
}

@article{Ozawa2019,
  title = {Topological photonics},
  author = {Ozawa, Tomoki and Price, Hannah M. and Amo, Alberto and Goldman, Nathan and Hafezi, Mohammad and Lu, Ling and Rechtsman, Mikael C. and Schuster, David and Simon, Jonathan and Zilberberg, Oded and Carusotto, Iacopo},
  journal = {Rev. Mod. Phys.},
  volume = {91},
  pages = {015006},
  year = {2019},
  doi = {10.1103/RevModPhys.91.015006}
}

@article{Shinjo2000,
  author = {Shinjo, T. and Okuno, T. and Hassdorf, R. and Shigeto, K. and Ono, T.},
  title = {Magnetic vortex core observation in circular dots of permalloy},
  journal = {Science},
  volume = {289},
  pages = {930--932},
  year = {2000},
  doi = {10.1126/science.289.5481.930}
}

@article{Wachowiak2002,
  author = {Wachowiak, A. and Wiebe, J. and Bode, M. and Pietzsch, O. and Morgenstern, M. and Wiesendanger, R.},
  title = {Direct observation of internal spin structure of magnetic vortex cores},
  journal = {Science},
  volume = {298},
  pages = {577--580},
  year = {2002},
  doi = {10.1126/science.1075302}
}

@article{NagaosaTokura2013,
  author = {Nagaosa, N. and Tokura, Y.},
  title = {Topological properties and dynamics of magnetic skyrmions},
  journal = {Nat. Nanotechnol.},
  volume = {8},
  pages = {899--911},
  year = {2013},
  doi = {10.1038/nnano.2013.243}
}

@misc{Code,
  author       = {{\DJ}or{\dj}evi{\'c}, Du{\v s}an and Molina, Fabi{\'a}n and Juri{\v c}i{\'c}, Vladimir},
  title        = {Code for  ``{W}ormhole {G}eometry from a {M}agnetic {V}ortex''},
  year         = {2026},
  howpublished = {\url{https://github.com/dusandjordjevic-ff/Vortex-Wormhole-.git}}
}

@article{Papanicolaou1991,
  title = {Dynamics of magnetic vortices},
  author = {Papanicolaou, N. and Tomaras, T. N.},
  journal = {Nucl. Phys. B},
  volume = {360},
  pages = {425},
  year = {1991},
  doi = {10.1016/0550-3213(91)90410-Y}
}

\end{document}


\title{Supplemental Material:\\ Wormhole Geometry from a Magnetic Vortex}

\author{Du\v{s}an \Dj or\dj evi\'{c}}
\affiliation{Faculty of Physics, University of Belgrade, Studentski Trg 12--16, 11000 Belgrade, Serbia}
\author{Fabi\'{a}n Molina}
\affiliation{Departamento de F\'{i}sica, Universidad T\'{e}cnica Federico Santa Mar\'{i}a, Casilla 110, Valpara\'{i}so, Chile}
\author{Vladimir Juri\v{c}i\'{c}}
\thanks{Corresponding author: vladimir.juricic@usm.cl}
\affiliation{Departamento de F\'{i}sica, Universidad T\'{e}cnica Federico Santa Mar\'{i}a, Casilla 110, Valpara\'{i}so, Chile}

\begin{abstract}
This Supplemental Material provides the technical details supporting the main
text. Section~\ref{sec:metric} derives the emergent metric from the quantum
geometry of the spin texture and establishes the wormhole form. Section
\ref{sec:roleofk} analyzes the role of the spiral wave vector and shows that the
throat is $\kk$-independent. Section~\ref{sec:lensing} derives the classical
lensing law and its reduction to an elliptic integral. Section~\ref{sec:berry}
derives the Berry-phase flux $\gamma=q/2$ and the geometric scalar potential.
Section~\ref{sec:cross} presents the partial-wave cross section, including the
analytic treatment of the Aharonov--Bohm contribution and the numerical
validation of the method. Section~\ref{sec:tb} gives the tight-binding
construction and the wave-packet simulation, together with a geodesic-level check
that the engineered lattice reproduces the Ellis deflection collapse.
\end{abstract}

\maketitle

\section{Emergent metric from the quantum geometry}\label{sec:metric}

\subsection{Quantum-metric correction}
A conduction electron Hund-coupled to a smooth local magnetization
$\spin(\rr)$, with coupling $J\gg E_F$, adiabatically follows the local spin
frame. Projecting the kinetic energy onto the locally aligned spinor produces,
in addition to an emergent gauge potential and scalar potential, a modification
of the effective mass tensor that is most compactly written as a metric. For a
unit-vector field $\spin$ the result of Ref.~\cite{Onishi:2025lde} is
\begin{equation}\label{eq:sm_qmetric}
    g_{ij}=\delta_{ij}+\frac{\hbar^2}{4mJ}\,
    \partial_i\spin\cdot\partial_j\spin\,,
\end{equation}
where the correction is the quantum metric (the symmetric part of the
quantum-geometric tensor) of the family of local spin coherent states. The
physical content is that the electron's effective hopping is reduced in
directions along which the spin frame rotates rapidly: a spatially varying
texture acts as a position-dependent effective mass, equivalently a curved
metric.

\subsection{Planar texture and reduction to a phase gradient}
We take the planar (easy-plane) texture
\begin{equation}
    \spin=\bigl(\cos\Phi,\,-\sin\Phi,\,0\bigr),\qquad
    \Phi(\rr)=\Psi(\rr)+\kk\cdot\rr,
\end{equation}
with $\Psi=\arg[(x-X)+i(y-Y)]$ the winding-one vortex phase. Since $\spin$ is
a planar unit vector, $\partial_i\spin\cdot\partial_j\spin
=\partial_i\Phi\,\partial_j\Phi$, and the metric correction collapses to an
outer product of the phase gradient:
\begin{equation}\label{eq:sm_outer}
    g_{ij}=\delta_{ij}+r_0^2\,\partial_i\Phi\,\partial_j\Phi,\qquad
    r_0^2=\frac{\hbar^2}{4mJ}.
\end{equation}
For a winding-$q$ vortex, $\Psi=q\phi$ and
$\partial_i\Psi=q\,(\hat{\mathbf{e}}_\phi)_i/r$, so
$\partial_i\Psi\,\partial_j\Psi=q^2(\hat{\mathbf{e}}_\phi)_i
(\hat{\mathbf{e}}_\phi)_j/r^2$.

\subsection{Wormhole form and Ricci scalar}
Setting $\kk=0$, the correction is purely azimuthal, and in polar coordinates
\begin{equation}\label{eq:sm_polar}
    \diff l^2=\diff r^2+\bigl(r^2+a_q^2\bigr)\diff\phi^2,\qquad a_q=|q|r_0.
\end{equation}
The substitution $\rho=\sqrt{r^2+a_q^2}$, with $\diff\rho=r\,\diff r/\rho$ and
$r^2=\rho^2-a_q^2$, gives $\diff r^2=\rho^2\,\diff\rho^2/(\rho^2-a_q^2)
=\diff\rho^2/(1-a_q^2/\rho^2)$, hence
\begin{equation}\label{eq:sm_ellis}
    \diff l^2=\frac{\diff\rho^2}{1-a_q^2/\rho^2}+\rho^2\diff\phi^2,
\end{equation}
the equatorial section of the Ellis wormhole with shape function
$b(\rho)=a_q^2/\rho$ and zero redshift. The areal radius obeys $\rho\ge a_q$,
with equality at the throat; the proper circumference $2\pi\rho$ is minimized
there at $2\pi a_q$.

The Gaussian (and, in $2{+}1$ dimensions with a static metric, Ricci) curvature
of Eq.~\eqref{eq:sm_polar} is computed from $\diff l^2=\diff r^2+h(r)^2
\diff\phi^2$ with $h=\sqrt{r^2+a_q^2}$ via $R=-2h''/h$. A direct computation
gives
\begin{equation}\label{eq:sm_ricci}
    R=-\frac{2a_q^2}{(r^2+a_q^2)^2}
    =-\frac{8\hbar^2 mJ\,q^2}{(\hbar^2 q^2+4mJr^2)^2}\,,
\end{equation}
{reproducing \rc{Eq.~(9)} of the main text.} The curvature is negative
everywhere, peaks at the throat ($r=0$) with $R=-2/a_q^2$, and decays as
$r^{-4}$, confirming asymptotic flatness. {\color{black}
The metric in Eqs.~\eqref{eq:sm_polar}--\eqref{eq:sm_ricci} is the leading term of the strong-exchange projection. Following Ref.~\cite{Onishi:2025lde}, define the exchange length $\ell_C=\hbar/\sqrt{mJ}$, the local texture scale $\lambda_s=|\nabla\spin|^{-1}$, and the electronic wavelength $\lambda_F\sim1/\kappa$. The projected Hamiltonian requires $J\gg E$ together with $\ell_C^2\ll\lambda_s\lambda_F$, while the local metric form additionally assumes $\lambda_s\gg\lambda_F$. For a winding-$q$ vortex, $\lambda_s\sim r/|q|$, so these conditions are parametrically controlled in the exterior and fail first near the microscopic core. The ratio $r_0^2/r^2$ diagnoses the size of the metric correction itself, but is not by itself the full adiabatic control parameter. The core therefore supplies the ultraviolet cutoff of the continuum Ellis geometry.
}

\section{Role of the spiral wave vector}\label{sec:roleofk}

Retaining $\kk\neq0$, the phase gradient is
$\partial_i\Phi=\partial_i\Psi+k_i$, so the metric correction
\eqref{eq:sm_outer} expands as
\begin{equation}\label{eq:sm_kdecomp}
    g_{ij}-\delta_{ij}=r_0^2\bigl(
    \partial_i\Psi\,\partial_j\Psi
    +k_i\partial_j\Psi+k_j\partial_i\Psi
    +k_ik_j\bigr).
\end{equation}
The trace, which controls the diagonal deformation, is
\begin{equation}\label{eq:sm_ktrace}
    g_{ii}-2=r_0^2\Bigl(\frac{1}{r^2}
    {\color{black}+\frac{2(-k_x\sin\alpha+k_y\cos\alpha)}{r}}
    +k^2\Bigr),
\end{equation}
where $\alpha$ is the azimuthal angle about the core. The three terms organize
by powers of $\kk$:
\begin{itemize}
\item $\mathcal{O}(k^0)$: the vortex term $r_0^2/r^2$, isotropic, responsible
for the throat;
\item $\mathcal{O}(k^1)$: the cross term $\propto 1/r$, \emph{anisotropic}
(angle-dependent), which vanishes upon averaging over $\alpha$;
\item $\mathcal{O}(k^2)$: the constant $r_0^2 k^2$, a uniform rescaling of the
asymptotic metric removable by a linear coordinate change.
\end{itemize}
Two conclusions follow. (i) The throat term is exactly independent of $\kk$, so
the wormhole throat and the angle-averaged geometry are properties of the
vortex winding alone. (ii) The cross term decays as $1/r$, more slowly than the
$1/r^2$ throat term, and therefore dominates the geometry at large distance: the
spiral reshapes the \emph{far-field} lens into a direction-dependent one
without altering the throat. A nonzero \(\kk\) thus adds anisotropic lensing without modifying the Ellis
throat scale. Here we set \(\kk=0\) to isolate the universal vortex geometry.

\section{Classical lensing law}\label{sec:lensing}

\subsection{Geodesic equation and deflection integral}
The spatial metric $\diff l^2=\diff r^2+f(r)^2\diff\phi^2$ with
$f(r)=\sqrt{r^2+a_q^2}$ has the geodesic Lagrangian
$\mathcal{L}=\half(\dot r^2+f^2\dot\phi^2)$. The cyclic coordinate $\phi$ gives
the conserved angular momentum $b=f^2\dot\phi$ (the impact parameter for a {trajectory}
of unit asymptotic speed), and the unit-speed condition $\dot r^2+f^2\dot\phi^2
=1$ becomes
\begin{equation}\label{eq:sm_geo}
    \dot r^2+\frac{b^2}{r^2+a_q^2}=1.
\end{equation}
The radial turning point is $r_{\min}=\sqrt{b^2-a_q^2}$ (real for $b>a_q$). The
angle swept from the turning point to infinity is
\begin{equation}\label{eq:sm_phiint}
    \varphi_\infty(b)=\int_{r_{\min}}^{\infty}
    \frac{b\,\diff r}{\sqrt{r^2+a_q^2}\,\sqrt{r^2+a_q^2-b^2}}.
\end{equation}

\subsection{Reduction to a complete elliptic integral}
Substituting $r=\sqrt{b^2-a_q^2}\,\cosh u$ maps $r_{\min}\to(u=0)$ and
$\infty\to(u\to\infty)$; with $t=\tanh u$ one obtains
\begin{equation}\label{eq:sm_elliptic}
    \varphi_\infty(b)=\int_0^1\frac{\diff t}{\sqrt{1-t^2}\,
    \sqrt{1-(a_q/b)^2\,t^2}}
    =K_{\mathrm{ell}}\!\left(\frac{a_q}{b}\right),
\end{equation}
where $K_{\mathrm{ell}}(x)=\int_0^{\pi/2}\diff\theta/\sqrt{1-x^2\sin^2\theta}$.
The total deflection is $\Theta(b)=2\varphi_\infty(b)-\pi
=2K_{\mathrm{ell}}(a_q/b)-\pi$. We verified Eq.~\eqref{eq:sm_elliptic}
numerically against direct integration of \eqref{eq:sm_phiint} (agreement to
$10^{-9}$) and against geodesic integration of the Hamilton equations (agreement
to $<3\times10^{-3}$ rad over $b/a_q\in[1.5,5]$).

\subsection{Limits}
For $b\to a_q^+$, $K_{\mathrm{ell}}(x)\to-\half\ln(1-x^2)+\ln 4$ as $x\to1^-$,
so $\Theta$ diverges logarithmically: grazing \rc{geodesics} wind around the throat. For
$b\gg a_q$, $K_{\mathrm{ell}}(x)=\tfrac{\pi}{2}(1+\tfrac{x^2}{4}
+\tfrac{9x^4}{64}+\cdots)$, giving
\begin{equation}\label{eq:sm_weak}
    \Theta(b)\simeq\frac{\pi a_q^2}{4b^2}
    =\frac{\pi q^2\hbar^2}{16mJ\,b^2},
\end{equation}
the universal $q^2/J$, $1/b^2$ far-field tail. Since $a_q$ is the only scale,
$\Theta$ is a function of $b/a_q$ alone, which is the data collapse of
Fig.~2 of the main text.

\section{Berry-phase flux and geometric scalar potential}\label{sec:berry}

\subsection{Emergent flux $\gamma=q/2$}
Adiabatic projection of the spin-$\tfrac12$ electron onto the local
quantization axis $\hat\spin(\rr)$ generates a $U(1)$ Berry connection
$\mathcal{A}=\langle u_{\hat\spin}|i\,\diff|u_{\hat\spin}\rangle$. For a spin
coherent state pointing along $\hat\spin$ with polar angles
$(\Theta_S,\Phi_S)$,
\begin{equation}\label{eq:sm_berry}
    \mathcal{A}=\half\,(1-\cos\Theta_S)\,\diff\Phi_S,
\end{equation}
the connection of a unit monopole. For the easy-plane vortex,
$\Theta_S=\pi/2$ and $\Phi_S=q\phi$, so
$\mathcal{A}=\half\,q\,\diff\phi$ and the flux enclosed in a loop around the
core is
\begin{equation}\label{eq:sm_flux}
    \oint\mathcal{A}=\half q\oint\diff\phi=\pi q,
\end{equation}
i.e.\ \emph{half} the naive $2\pi q$ winding flux---the planar (easy-plane)
texture sits on the equator of the spin sphere, halving the solid angle. This is
the Aharonov--Casher phase of a spin transported around a textured field. In the
scattering problem it enters as an Aharonov--Bohm flux $\gamma=q/2$, shifting
the angular momentum $\ell\to\ell-\gamma$. {\color{black}The overall sign of $\gamma$ depends on the orientation and spinor gauge convention; only $\gamma$ modulo an integer enters the cross section, so the parity fingerprint below is convention independent.} The elementary vortex $q=1$ thus sits at half-integer flux $\gamma=\half$, producing the strongest nontrivial Aharonov--Bohm interference pattern.

\subsection{Geometric scalar potential}
The same projection produces a scalar potential from the gradient of the local
frame. For the easy-plane spin-$\tfrac12$ texture it is
\begin{equation}\label{eq:sm_ugeom}
    U_{\mathrm{geom}}(r)=\frac{\hbar^2}{8m}(\nabla\Psi)^2
    =\frac{\hbar^2 q^2}{8mr^2}\,,
\end{equation}
a repulsive inverse-square potential. It is long ranged ($\sim r^{-2}$, the same
power as the centrifugal term) and is distinct from the finite-$J$ geometric
lensing: it survives even in the $J\to\infty$ limit where the metric becomes
flat. In the radial equation it enters as
$2mU_{\mathrm{geom}}/\hbar^2=q^2/(4r^2)$ in units $a_q=1$, i.e.\ an effective
angular-momentum shift $(\ell-\gamma)^2\to(\ell-\gamma)^2+q^2/4$ at large $r$.
{\color{black}At large $|\ell-\gamma|$ the scalar term produces
\[
\delta_\ell^{\rm scalar}\simeq-\frac{\pi q^2}{16|\ell-\gamma|},
\]
whereas the metric contribution is $\delta_\ell^{\rm metric}\simeq\pi(\kappa a_q)^2/[8|\ell-\gamma|]$. Since $\kappa a_q=|q|\sqrt{E/(2J)}$, their magnitude ratio scales as $|\delta^{\rm metric}/\delta^{\rm scalar}|\sim E/J$. Thus the scalar term dominates the asymptotic quantum-scattering phase in the controlled $J\gg E$ magnetic regime, while the geodesic deflection remains a clean probe of the metric.}

{\color{black}
\section{Partial-wave cross section}\label{sec:cross}

\subsection{Radial equation and diagnostic decomposition}

The projected vortex ties together the Ellis scale $a_q=|q|r_0$, the Berry flux $\gamma=q/2$, and the scalar-potential coefficient $C=q^2/4$, defined through $2mU_{\mathrm{geom}}/\hbar^2=C/r^2$. To diagnose their distinct roles we treat the throat radius $a$, $\gamma$, and $C$ as independent parameters of the effective scattering Hamiltonian. In units $a=1$,
\begin{equation}\label{eq:sm_radial}
R_\ell''+\frac{r}{r^2+1}R_\ell'+\left[\kappa^2-\frac{(\ell-\gamma)^2}{r^2+1}-\frac{C}{r^2}\right]R_\ell=0.
\end{equation}
The microscopic vortex values are $a=a_q$, $\gamma=q/2$, and $C=q^2/4$. The geometry-only diagnostic keeps $a\neq0$ while setting $\gamma=C=0$; it is not a physical $q=0$ vortex, for which $a_q=0$.

{\color{black}
At the throat, the finite Ellis term makes $(\ell-\gamma)^2/(r^2+a^2)$ regular; only the scalar term is singular. Near the origin, the regular solution has the power-law form
\(R_\ell\sim r^s\), with \(s(s-1)=C\), and hence
\(s=\tfrac12[1+\sqrt{1+4C}]\) for \(C>0\). For the geometry-only diagnostic, $C=0$, we choose the even regular continuation, $R_\ell'(0)=0$; other microscopic core boundary conditions can modify low partial waves, so near-throat cross section is nonuniversal.

The asymptotic $1/r^2$
 tail is treated analytically, avoiding  matching to integer-order  Bessel functions at finite radius. Defining
\begin{equation}\label{eq:sm_nuinf}
\nu_\infty=\sqrt{(\ell-\gamma)^2+C},
\end{equation}
we integrate Eq.~\eqref{eq:sm_radial} outward from $r_{\min}=10^{-3}$ and match the logarithmic derivative $L=R_\ell'/R_\ell$ at $R$ to Bessel functions of order $\nu_\infty$,
\begin{equation}\label{eq:sm_match}
\tan\delta_\ell^{\rm th}=\frac{\kappa J_{\nu_\infty}'(\kappa R)-LJ_{\nu_\infty}(\kappa R)}{\kappa Y_{\nu_\infty}'(\kappa R)-LY_{\nu_\infty}(\kappa R)}.
\end{equation}
The total phase is then
\begin{equation}\label{eq:sm_totalphase}
\delta_\ell=\frac{\pi}{2}\bigl(|\ell|-\nu_\infty\bigr)+\delta_\ell^{\rm th}.
\end{equation}
For $C=0$ and $a\to0$, the first term reduces to the Aharonov--Bohm phase. This procedure removes the finite-matching-radius error associated with the long-range AB/scalar tail.

The large-$|\ell|$ metric contribution follows from the weak-field lensing law,
\begin{equation}\label{eq:sm_phaseshift}
\delta_\ell^{\rm metric}\simeq\frac{\pi(\kappa a)^2}{8\nu_\infty}\simeq\frac{\pi(\kappa a)^2}{8|\ell-\gamma|},
\end{equation}
while the scalar contribution relative to the pure AB phase is
\begin{equation}\label{eq:sm_scalarphase}
\delta_\ell^{\rm scalar}=\frac{\pi}{2}\left[|\ell-\gamma|-\nu_\infty\right]\simeq-\frac{\pi C}{4|\ell-\gamma|}.
\end{equation}
Both tails are algebraic. Accordingly, the numerical amplitude below uses exact phase shifts through $|\ell|\le L_{\max}$ and attaches Eqs.~\eqref{eq:sm_phaseshift}--\eqref{eq:sm_scalarphase} at larger $|\ell|$. For the geometry-only benchmark at $\kappa a=2$, the extracted phases approach Eq.~\eqref{eq:sm_phaseshift}, as shown in Table~\ref{tab:sm}.
}
}

\begin{table}[t]
\caption{{\color{black}Numerical phase shift $\delta_\ell$ for the diagnostic geometry-only problem, with $a=1$, $\gamma=C=0$, and $\kappa a=2$, compared with $\pi(\kappa a)^2/(8|\ell|)$.}}
\label{tab:sm}
\begin{ruledtabular}
\begin{tabular}{cccc}
$\ell$ & $\delta_\ell^{\,\mathrm{num}}$ & $\delta_\ell^{\,\mathrm{sc}}$ & ratio \\
\hline
$8$  & $0.1987$ & $0.1963$ & $1.012$ \\
$10$ & $0.1583$ & $0.1571$ & $1.008$ \\
$14$ & $0.1126$ & $0.1122$ & $1.004$ \\
$20$ & $0.0787$ & $0.0785$ & $1.002$ \\
$30$ & $0.0524$ & $0.0524$ & $1.000$ \\
\end{tabular}
\end{ruledtabular}
\end{table}

\begin{figure}[t]
\centering
\includegraphics[width=0.92\columnwidth]{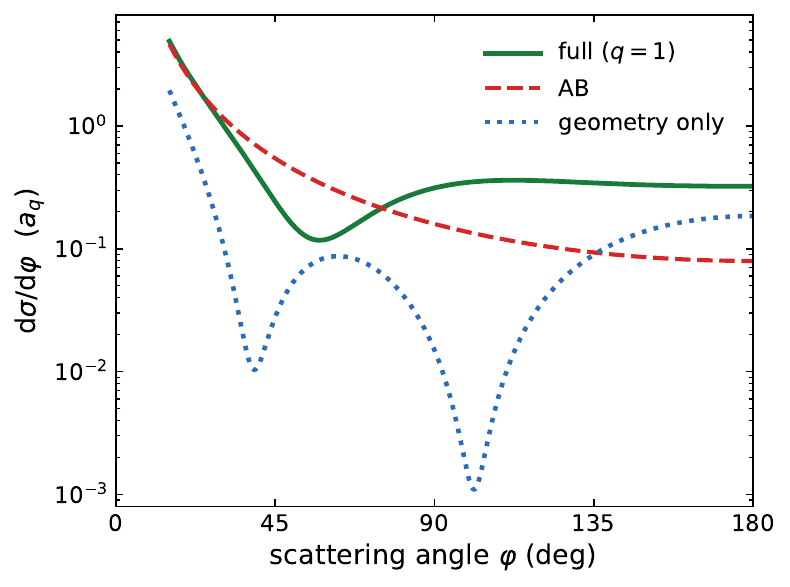}
\caption{{\color{black}Diagnostic differential cross section of the effective finite-throat problem at $\kappa a=2$. The geometry-only curve (dotted blue) sets $\gamma=C=0$; the AB curve (red dashed) is the closed-form half-flux response at $\gamma=1/2$; and the combined curve (green solid) uses $\gamma=1/2$, $C=1/4$. The example illustrates coherent interference between the universal AB background and finite-throat/scalar scattering. It is not a controlled $q=1$ magnetic material parameter point in the $J\gg E$ limit.}}
\label{fig:sm_xsec}
\end{figure}

{\color{black}
\subsection{Aharonov--Bohm resummation}
The differential cross section is $\diff\sigma/\diff\phi=|f(\phi)|^2$, with
\begin{equation}\label{eq:sm_amp}
f(\phi)=\sqrt{\frac{1}{2\pi\kappa}}e^{i\pi/4}\sum_\ell\left(e^{2i\delta_\ell}-1\right)e^{i\ell\phi}.
\end{equation}
For the pure AB part, $\delta_\ell^{\rm AB}=\frac{\pi}{2}(|\ell|-|\ell-\gamma|)$, the series is conditionally convergent. We therefore split
\begin{equation}\label{eq:sm_split}
e^{2i\delta_\ell}-1=\left(e^{2i\delta_\ell^{\rm AB}}-1\right)+\left(e^{2i\delta_\ell}-e^{2i\delta_\ell^{\rm AB}}\right).
\end{equation}
The first term is Abel resummed analytically, giving
\[
\left.\frac{\diff\sigma}{\diff\phi}\right|_{\rm AB}=\frac{\sin^2(\pi\gamma)}{2\pi\kappa\sin^2(\phi/2)}.
\]
{\color{black}The second term contains the finite-throat and scalar corrections. Since it retains the algebraic $1/|\ell|$ tails in Eqs.~\eqref{eq:sm_phaseshift}--\eqref{eq:sm_scalarphase}, the code computes exact phases through $L_{\max}=30$ and continues the analytic asymptotic remainder to $L_{\rm tail}=5000$ with an Abel regulator. Moving the exact/asymptotic join from $|\ell|=20$ to $30$ changes the full cross section by about $0.1\%$ away from diffraction zeros. The pure AB amplitude is evaluated directly in its exact Abel-resummed closed form.}

\subsection{Benchmarks and diagnostic scattering structure}
The construction passes three checks: it reproduces the exact pure-AB cross section; the geometry-only phases approach Eq.~\eqref{eq:sm_phaseshift} (Table~\ref{tab:sm}); and the AB forward singularity disappears for integer $\gamma$, as required by flux periodicity.

Figure~\ref{fig:sm_xsec} uses $\kappa a=2$ to display the structure of the effective Ellis scattering Hamiltonian. The geometry-only curve sets $\gamma=C=0$; the AB curve is the closed-form half-flux result; and the combined diagnostic uses $\gamma=1/2$, $C=1/4$ at the same finite throat radius. Their amplitudes are added coherently. This parameter point characterizes the quantum-scattering structure of the effective Ellis geometry. For a microscopic Hund-coupled vortex, where $a=a_q$, one has
\begin{equation}\label{eq:sm_kappaaq}
\kappa a_q=|q|\sqrt{\frac{E}{2J}},
\end{equation}
so $J\gg E$ implies $\kappa a_q\ll1$ for $q=1$. Figure~\ref{fig:sm_xsec} therefore serves as a diagnostic of the effective finite-throat quantum problem, while the controlled magnetic signature emphasized in the main text is the far-field geodesic deflection. The diffraction minima are correspondingly interpreted as features of the ideal effective Ellis Hamiltonian; their quantitative robustness to a microscopic core is not assumed.
}

\section{Tight-binding construction and wave-packet simulation}\label{sec:tb}

\subsection{Curved-space Dirac Hamiltonian}
With the orthonormal frame $e^1=\diff r$, $e^2=f(r)\diff\phi$,
$f=\sqrt{r^2+a_q^2}$, the massless Dirac Hamiltonian on the wormhole spatial
metric is
\begin{equation}\label{eq:sm_dirac}
    H=-i\hbar v_F\Bigl[\sigma^r\Bigl(\partial_r+\frac{f'}{2f}\Bigr)
    +\sigma^\phi\frac{1}{f}\partial_\phi\Bigr]+V(r),
\end{equation}
where $f'/2f=r/[2(r^2+a_q^2)]$ is the spin connection ensuring Hermiticity with
respect to the curved measure $f\,\diff r\,\diff\phi$. The inverse vielbein has
$e^r_1=1$ and $e^\phi_2=1/f(r)$, so the angular group velocity is reduced,
\begin{equation}\label{eq:sm_vel}
    \frac{v_\phi(r)}{v_F}=\frac{r}{f(r)}=\frac{r}{\sqrt{r^2+a_q^2}}
    \simeq1-\frac{a_q^2}{2r^2}\quad(r\gg a_q),
\end{equation}
while the radial velocity is unchanged.

\subsection{Engineered hoppings}
On a honeycomb lattice with nearest-neighbor bonds $\bm\delta_n$ ($n=1,2,3$),
the long-wavelength Dirac velocity is set by the bond-resolved hoppings.
Reducing the tangential bonds near the core according to
\begin{equation}\label{eq:sm_hop}
    t_n(\mathbf{R})=t_0\Bigl[1-\frac{a_q^2}{2(r^2+\xi^2)}
    (\hat{\bm\delta}_n\cdot\hat{\mathbf{e}}_\phi)^2\Bigr],\quad r=|\mathbf{R}|,
\end{equation}
with $\xi$ a core regularization, realizes Eq.~\eqref{eq:sm_vel} at leading
order. More generally, the three hoppings can be tuned to match the full local
vielbein $v_F e^i_a(r)$. {\color{black}In artificial-graphene, photonic, or topolectric honeycomb platforms, the bond-resolved hoppings and on-site terms are independently controllable. The simple modulation in Eq.~\eqref{eq:sm_hop} reproduces the target velocity form to leading order in the exterior but also generates a valley-odd pseudogauge component. In the wave-packet calculation below we remove this component by symmetrizing the $K$ and $K'$ responses, thereby isolating the valley-even metric dynamics. The independently controllable on-site term is set to $V(\mathbf R)=0$ in this calculation, so the propagation test isolates the programmed vielbein rather than the scalar potential.}

\begin{figure*}[t!]
    \centering
\includegraphics[width=\textwidth]{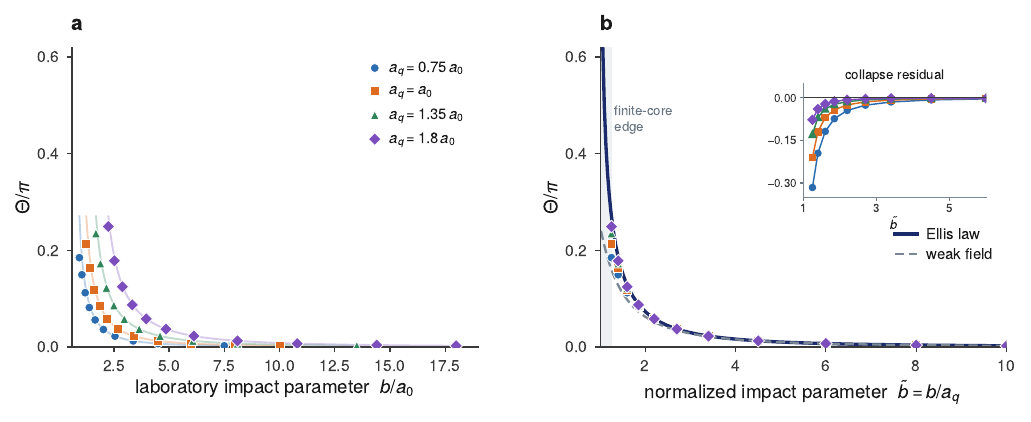}
    \caption{Core-induced distortion of the Ellis deflection law.
(a)~Deflection of geodesics of the core-regularized engineered metric, in
laboratory units $b/a_0$, for several throat radii $a_q$ at fixed  core
size $\xi$, where $a_0$ is a fixed reference length; the curves differ because
$a_q$ differs. (b)~The same data against $\tilde b=b/a_q$, compared with the
continuum Ellis law~\eqref{eq:sm_elliptic} (solid) and its weak-field tail
$\pi/(4\tilde b^2)$ (dashed). For an ideal vielbein the curves would collapse
exactly, since $a_q$ is then the only scale; the fixed core introduces the
second ratio $\xi/a_q$ and breaks the collapse near the throat. The grey band
marks the finite-core region. Inset: the residual, a direct measure of the core
truncation---of order unity for $\tilde b\lesssim2$, where a fixed 
core affects smaller throats more strongly, and below $1\%$ for 
$\tilde b\simeq6$.}
    \label{fig:sm_collapse}
\end{figure*}

\subsection{Wave-packet propagation}
{\color{black}
We build the tight-binding Hamiltonian on a honeycomb patch ({\color{black}$\simeq1.6\times10^5$} sites) with the hoppings~\eqref{eq:sm_hop}. One Ellis radius is represented by {\color{black}$a_q=19$} lattice constants, with core cutoff $\xi^2=0.12a_q^2$, packet width $\sigma\simeq0.9a_q$, impact parameter $b=1.2a_q$, and launch position $x_0=-10a_q$. These ratios correspond directly to the code convention $\mathrm{AQ}=1$ and {\color{black}$\mathrm{LU}=19$}.

For each valley we prepare the positive-energy Bloch eigenstate whose group velocity points along $+x$; this removes the spurious transverse velocity that an arbitrary equal-weight sublattice spinor would produce. The bond modulation nevertheless carries a valley-odd pseudogauge response. We therefore propagate otherwise identical $K$ and $K'$ packets and average their densities and centroids. This valley symmetrization cancels the pseudogauge deflection while preserving the valley-even metric response.

The reference ray is integrated in the same core-regularized metric represented by the hopping profile,
\begin{equation}\label{eq:sm_coremetric_emulator}
\diff l^2=\diff r^2+h(r)^2\diff\phi^2,\qquad h(r)^2=r^2+a_q^2\frac{r^2}{r^2+\xi^2},
\end{equation}
so the lattice and continuum comparisons use the same ultraviolet regularization. In the exterior tracking window, the valley-symmetrized centroid follows this geodesic {\color{black}over $8.6a_q$ of path} to $|\delta_\perp|<0.02a_q$. {\color{black}The simulation is restricted to this controlled exterior window; beyond the tracking interval the continuum geodesic is shown only as an extrapolation, and no quantitative near-core wave-packet claim is made. Supplemental Movie~1 gives the corresponding evolution.

As a null benchmark, we repeat the calculation with the hopping modulation switched off. With the same finite simulation window, the flat control remains straight to within {\color{black}$2.4\times10^{-3}a_{\rm ref}$ over $13.6a_{\rm ref}$}, i.e.\ over a longer path than the vortex tracking window, where {\color{black}$a_{\rm ref}=19$} lattice constants is the reference length used for the vortex run (Supplemental Movie~2). This control tests lattice discretization and packet preparation independently of the engineered metric.}
}

\subsection{Numerical methods and reproducibility}
{\color{black}
Geodesics are integrated with an adaptive Runge--Kutta scheme (\texttt{scipy.integrate.solve\_ivp}, \texttt{rtol}$=10^{-8}$) using the metric in Eq.~\eqref{eq:sm_coremetric_emulator} for the wave-packet comparison. Phase shifts use logarithmic-derivative matching at $R=160a$ after analytically separating the asymptotic Bessel order $\nu_\infty$ in Eq.~\eqref{eq:sm_nuinf}. The pure-AB amplitude is evaluated in its exact Abel-resummed closed form; the finite-throat correction uses exact phase shifts through $L_{\max}=30$ and the Abel-regularized analytic tail through $L_{\rm tail}=5000$. The script reports a $\simeq0.1\%$ change away from diffraction zeros when the exact/asymptotic join is moved from $|\ell|=20$ to $30$. Tight-binding evolution uses sparse \texttt{expm\_multiply} in Krylov blocks, which is algebraically identical to stepwise propagation but substantially reduces numerical overhead. {\color{black}The matching scheme of Eqs.~\eqref{eq:sm_nuinf}--\eqref{eq:sm_totalphase}, the cutoffs $L_{\max}$ and $L_{\rm tail}$, and the convergence check quoted above are those implemented in the accompanying script; Table~\ref{tab:sm} and Fig.~\ref{fig:sm_xsec} are generated by that script.}
}

\subsection{Core-induced distortion of the lensing law}
\mc{The engineered hoppings realize the target vielbein only outside the
microscopic core, and it is useful to know how far this truncation propagates
into the lensing observable. We therefore quantify the deflection of the
core-regularized engineered metric $\diff l^2=\diff r^2+h(r)^2\diff\phi^2$, with
$h(r)^2=r^2+a_q^2\,g(r)$, by integrating
$\Theta(b)=2\!\int_{r_{\min}}^{\infty}\! b\,\diff r/[\,h\sqrt{h^2-b^2}\,]-\pi$
independently for several throat radii $a_q$, in laboratory units. For the
ideal vielbein, $g(r)=1$, the metric has $a_q$ as its only scale, so $\Theta$
depends on $b/a_q$ alone and reproduces the closed-form result
\eqref{eq:sm_elliptic}; the curves then collapse exactly. A lattice core of
fixed absolute size softens the angular term, $g(r)=1/(1+\xi^2/r^2)$, and
introduces the second dimensionless ratio $\xi/a_q$, so the collapse is no
longer exact [Fig.~\ref{fig:sm_collapse}]. The residual is therefore a direct
measure of the core truncation: it is of order unity for $b\lesssim2\,a_q$,
where a fixed absolute core affects smaller throats more strongly, and falls
below $1\%$ by $b\simeq6\,a_q$. Two consequences used in the main text  follow directly: in the far field the deflection is
core-insensitive, so the weak-field tail~\eqref{eq:sm_weak} applies without
detailed knowledge of $\xi$, whereas resolving the near-throat geometry requires
$\xi\ll a_q$. Both are ray-level statements, bounding the distortion from the
core but not from packet spreading, which Sec.~S6\,C quantifies separately.}